# Probabilistic-Bits based on Ferroelectric Field-Effect Transistors for Stochastic Computing

*Sheng Luo\*, Yihan He, Baofang Cai, Xiao Gong, Gengchiau Liang\**

Department of Electrical and Computer Engineering, National University of Singapore, Singapore 117576

**Abstract:**

A probabilistic-bit (p-bit) is the fundamental building block in the circuit network of stochastic computing, and it could produce a continuous random bitstream with tunable probability. Utilizing the stochasticity in few-domain ferroelectric material (FE), we propose for the first time, the p-bits based on ferroelectric FET. The stochasticity of the FE p-bits stems from the thermal noise-induced lattice vibration, which renders dipole fluctuations and is tunable by an external electric field. The impact of several key FE parameters on p-bits' stochasticity is evaluated, where the domain properties are revealed to play crucial roles. Furthermore, the integer factorization based on FE p-bits circuit network is performed to verify its functionality, and the accuracy is found to depend on FE p-bits' stochasticity. The proposed FE p-bits possess the advantages of both extremely low hardware cost and the compatibility with CMOS-technology, rendering it a promising candidate for stochastic computing applications.

# Introduction

The classical computers based on the Turing machine and von Neumann architecture are the centrepieces of today's information society. In standard binary computing circuitry, the information is stored and computed with deterministic stable units, such as the on/off states of a transistor. However, there are many computational challenging problems that the classical computers could not address efficiently, such as the combinatorial optimization problem (COP) [1], integer factorization (IF) [2], Bayesian inference [3], etc. These problems usually have a large subset size, and the computational complexity exponentially increases under classical unidirectional circuitry. Stochastic computing (SC), therefore, is considered as an alternative solution [4] to overcome these technical challenges. Compared with the deterministic "bits" in traditional binary circuits, SC makes use of unstable units called the probabilistic bits (p-bits) [4] to serve as the building blocks, and it could fluctuate rapidly between the binary states 0 and 1 to produce a continuous random telegraphic signal. More importantly, different from the other devices with random characteristics, the behaviors of p-bits possess tunable probability with sigmodal relations against external control signals. The probabilistic nature of the p-bits not only enables some low-cost implementations of arithmetic operations compared to the standard logic elements [5-6], but also offers a high error tolerance computational method to counter the transient or process-variations-induced errors [7]. Consequently, the SC based on p-bit networks has been recognized as a useful computational solution in the system where conditions of small size, low power, and soft-error tolerance are required [7]. It serves as an important solution to image recognition [8], neural networks [9], and machine learning [10].

Several designs of p-bits have been proposed based on different probability generation and adjustment mechanisms. In recent years, it has been demonstrated that the CMOS-based p-bits could utilize the fully-CMOS modules, such as the linear feedback shift-register (LFSR) [11-12] and field programmable gate array (FPGA) [13], to introduce the pseudo-random bitstreams for SC applications. Although the CMOS-based p-bits are inherently compatible with the CMOS technology nodes, they have high device density (usually up to a few thousand transistors) and energy consumption [14]. In contrast, another proposed p-bit design based on magnetic tunnel junction (MTJ) [2,14] has significantly lower hardware cost as the p-bit is based on either $1T$ (transistor)+$1M$ (MTJ) or $1R$ (resistor)+$1M$ structure. The relatively low device density also curtails the energy consumption to a few fJ per p-bit [14]. Furthermore, different from the pseudo noise implemented in CMOS-based p-bits as the source of randomness, MTJ p-bits utilize the true random thermal noise-induced stochasticity in the ferromagnetic material (FM). Though the MTJ p-bits possess the advantages of extremely low hardware cost and power consumption, the complicated scaling process and compatibility with CMOS technology nodes still pose significant challenges for its applications in large-scale p-bits networks.

Similar to the stochasticity in the FM system, the dynamic behaviours of the ferroelectric material (FE) are random as well, enabling it as a prospective candidate for p-bits design. One commonly utilized random feature of the FE system could be observed in the repeated switching process under a series of pulses [15-20]. Differences in the switching threshold among the grains induce variations in each switching process performed consecutively. Based on this characteristic, the true random number generator [15] and stochastic spiking neuron [19] are proposed based on the ferroelectric FET (FeFET) and ferroelectric tunnel junction (FTJ). The nucleation limited switching (NLS) [18] with the Monte Carlo scheme is implemented to model this random switching process. Nevertheless, this random process is not suitable for p-bit's design, as it does not have the mechanism for the adjustable probability of its polarization. Another source of the randomness in the FE system stems from temperature-dependent effects. The thermal noise-induced lattice vibration, for example, is an important mechanism in the nucleation creation and the domain's growth process in FE dynamics, especially around the defects and grain boundaries [21-22]. This mechanism could assist the FE domains to overcome the energy barriers and induce the real-time random fluctuations of the FE dipoles [21-30], different from the randomness in repeated switching. The fluctuating dipoles play a crucial role in the FE dynamic properties, such as the dipole reversal switching [23-28,30], dielectric response [29], and domain motions [24,30]. The lattice vibrations induced fluctuating dipoles could be characterized as random Brownian motions with an average fluctuating position in the phase-field model following a Boltzmann-type probability distribution [24-28]. The average position could be further adjusted by the external electric field, as shown in Fig. 1a. Through the FE-based nanodevice, such as the FeFET, the fluctuations of the dipoles could be further converted into the randomly fluctuating current with a tunable fluctuation range. With a designated voltage or current threshold, a mechanism of the controllable probability adjusted by the external electric field could be established.

In this work, for the first time, we propose the stochastic p-bit based on FeFET (Fig. 1b), utilizing the thermal noise-induced dipole fluctuation in a few-domain FE material. The fluctuation range of the FE dipole could be adjusted by the gate bias $V_g$ (Fig. 1c), and through the post-processing circuit modules (Fig. 1d), the analogue fluctuating-current-signal could be converted to the random binary bitstream under certain voltage threshold $V_{th}$. This allows the gate bias to adjust the probability of the FE p-bits output against the $V_{th}$, characterized by the probabilistic curve (p-curve) with sigmoidal relation (Fig. 1e). The impact of several key FE material parameters on stochasticity is analysed through the phase-field model, and the crucial roles of domain coupling and domain number are revealed in the suppression of stochasticity in the FE system. Finally, we verify the proposed p-bits' functionality in SC applications through the integer factorization (IF) carried out in FE p-bits based invertible logic (IL) circuit. The IF performance, such as the accuracy, demonstrates its dependency on the FE p-bits stochasticity and the importance of FE p-bits design for SC applications. The proposed FE-based p-bit is found to possess both advantages of extremely low hardware cost and compatibility with CMOS technological nodes, amending the drawbacks of both fully CMOS and MTJ-based p-bits.

## Results

### Thermal noise-induced stochasticity in FE

For the convex free energy landscape of the FE material, the spontaneous polarization (SP) locates in the global energy minima, designating the states where the FE system relaxes. The external electric field applied to the FE system changes the system's free energy and the global energy minima, and thus the SP position alters accordingly. The polarization transition rate (PTR), defined as $\left(\frac{dp}{dt}\right)$ in time-dependent Landau-Ginzburg (TDGL) equations (see Methods for phase field model details), reaches zero when the free energy of the FE system at energy minima, as shown in Fig. 2a. The insets show the free energy under electric field in two opposite directions, and the PTR shows the SP aligns with the direction of electric field (labelled as $A$ and $A'$) in accordance. The PTR could be shifted upward/downward by electric field, changing both the SP value and the transition speed of the polarization. Under this picture, the thermal noise-induced polarization shifting, shown as $+\eta$ and $-\eta$ in Fig. 2b, could either enhance or thwart the FE switching process, rendering higher/lower PTR (Point B or C). This coincides with the thermal noise-assisted polarization reversal switching observed in both experiments and simulations [23-28,30]. More importantly, the $\eta$ also alters the original SP position to either B' or C', dragging the polarization to either direction. Therefore, the continuously random $\eta$ could effectively create a real-time polarization fluctuation. To further illustrate the fluctuation of polarization, three different constant electric fields are applied to the FE system with different free energy profiles (Fig. 2c). The statistical distribution of polarization is extracted under the observation time up to 1 μs (including $10^4$ sampling points). As the applied field increases, the entire distribution shifts as well. Moreover, the increment of the electric field reduces the energy of the global minima (Fig. 2c). This leads to the reduction of the fluctuation range as the FE polarization fluctuates closer around the SP positions (Fig. 2d), reducing the probability of the FE system stays in the opposite polarization direction.

To further understand the stochastic mechanisms in FE system, the impact of several key FE material parameters on stochasticity is evaluated. Firstly, the impact of Landau coefficients ($\alpha_1$, $\alpha_{11}$ and $\alpha_{111}$) is evaluated under single domain assumption, with two different sets of coefficients, designated as FE-1 and FE-2 (parameters detailed in Methods). The corresponding free energy profiles are shown in Fig. 3a. The slope of the PTR, which could be approximated as the second order derivative of polarization potential energy $\Psi_{pol}$ of the FE system ($\frac{\partial^2 p}{\partial t \partial p} \propto -\frac{\partial^2 \Psi_{pol}}{\partial p^2}$), could be implemented to estimate the relation between Landau coefficients and fluctuation range. Under the same thermal noise-induced polarization shifting $\eta$, the steeper slope of PTR renders a smaller fluctuation range, as illustrated in the blue and red triangles in Fig. 3b. This could be further verified in Fig. 3c, as the statistic distributions of polarization demonstrate the relatively wider range of fluctuation for FE system with FE-1 Landau coefficients comparing those of the FE-2. The second-order derivative of the free energy is also related to the barrier height in the FE material, which demonstrates the relation between barrier height and dipole fluctuation range.

Another FE parameter evaluated is the domain coupling strength $G_0$, an important parameter characterizing the domain interaction. To evaluate its impact on FE stochasticity, a two-domain FE system is investigated. Each domain has a different set of Landau coefficients, which is either FE-1 or FE-2. Different domain wall profiles could be observed as domain interaction changes under the same external field (Fig. 3d), demonstrating the extent of SP alterations around the interface as $G_0$ changes. The stronger domain interaction reduces the differences in polarizations among the different domains. This could be reflected in the PTR of each domain under different $G_0$, as shown in Fig. 3e. Polarization of each domain fluctuates around each domain's SP, which is separated due to the Landau coefficient differences. It is noteworthy that the enhanced $G_0$ moves both domain's SP inwards, reducing the overall alteration range. This could further be verified in Fig. 3f by extracting the distribution in extensive observation time of 1 $\mu s$. The stronger domain interaction increases the overlapping of the polarization fluctuation range of the two domains, while the polarizations of the weaker coupling FE domain occupy a wider range.

**Stochasticity in the multidomain FeFET p-bits**

By clarifying the mechanism of the thermal noise-induced random polarization fluctuations in FE system, the stochasticity of the FeFET-based p-bits could be further evaluated (see Methods for device modeling scheme of FeFET). To approach the multi-domain FE system, the Landau coefficients and the viscosity coefficients are assumed to follow the Gaussian distribution in a multidomain system, with original parameters as mean values and deviations σ up to 20% [31]. This aims to approach the FE system with doping, grain boundary, heterojunction, etc., instead of the uniform parameters in the FE system [32-33]. An example of the *I-V* characteristics of FeFET p-bits is shown in Fig. 4a, in which the FE system consists of 12 domains. As the results of the stochasticity in the FE system, the output drain current $I_d$ fluctuates at each sample gate bias under constant gate bias for extensive observation time (up to few μs). The FeFET is first initialized by $V_{init}$ (inset of Fig. 4a). Then, the different gate bias ($V_{g1}$, $V_{g2}$, …) is applied to the system, and remains constant within the observation time. The *I-V* characteristics of the stochastic FeFET (blue dot lines) in Fig. 4a record all the output current within the observation time, demonstrating up to 2 orders of current differences in the fluctuation. The cyan dash lines are the *I-V* curves of the same FeFET in a forward/reverse sweep without the thermal noise. In the relatively large gate bias (~3 V), the range of fluctuation in current reduces as the FE system are fully switched, and minor capacitance variations could be induced from the dipole fluctuations. Therefore, both *I-V* curves reach similar output current. A zoom-in view of the statistic distribution of current at each sample gate bias is shown in Fig. 4b. The stochasticity in the FE system introduces a "broadening" in the FeFET's current distribution, with the fluctuation range shifted by different gate bias. The p-curve in Fig. 4c illustrates the gate bias adjustment of the FE p-bits' probability in surpassing the fixed voltage threshold $V_{th}$, configured by the inverters in Fig. 1c. Without the inclusion of the thermal noise, the drain current of FeFET is deterministic under certain gate bias, and it produces the output voltage $V_{out}$ (defined in Fig. 1c) as either 0 (below $V_{th}$) or 1 (above $V_{th}$) in p-curve as a step function (black dash line in Fig. 4c). In contrast, the broadening induced by thermal noise in *I-V* characteristics introduce a slope in the p-curve. The steepness could be further adjusted by temperature, as increasing temperature enhances the lattice vibration and the stochasticity of the p-bits (shown as 150 K and 300 K, respectively). To further illustrate the extraction of the probability at the sampled gate bias, the output voltage $V_{out}$ at points *A*, *B* and *C* are demonstrated in Fig. 4d. The stochasticity in a FE system enables the drain current to possess certain probability to reach above the $V_{th}$. The probability is defined as $p = \frac{N_{(Vout>Vth)}}{N_{total}}$, where $N_{(Vout>Vth)}$ is the total number of time frame as the p-bit's output is above the threshold, and the $N_{total}$ is the total sampling time. As the fluctuation range of the current could be shifted by the gate bias, different portions of time frames above the threshold are demonstrated. The p-curve of various sampled voltage within the operation window could be fitted with sigmoidal relations, as the solid blue and purple lines in Fig. 4c.

As the stochasticity of the FE system plays a crucial role in the probabilistic property of the p-bits, the stochasticity of FeFET is further evaluated regarding several aspects: the domain coupling, variations of the Landau coefficients, and the number of FE domains. Different domain coupling strengths ($G_0$) are implemented in a four-domain system to investigate its impact (Fig. 5a), and it demonstrates the suppression of the current fluctuation range as $G_0$ enhances. This derives from the mechanism of narrowed polarization fluctuation range as discussed in Fig. 3f, which suppresses the fluctuations of the capacitance of FE. The corresponding p-curves shown in Fig. 5b demonstrate a steeper slope (less stochastic) as the domain interaction enhances. The impact of the FE system with different deviations of Landau coefficients is also evaluated as σ increases to 40% and 60% (Fig. 5c). The enlarged current fluctuation range could be observed with increasing domain-to-domain variations. The increment of the deviations allows the domains with a wider polarization fluctuation range could be included, and thus enhance the overall stochasticity in the output current, reflected in the p-curve of the corresponding FeFET (Fig. 5d).

Furthermore, the stochasticity of FeFET p-bits with different domain numbers is evaluated (Fig. 6a). Both Landau coefficients and viscosity coefficients $\mu$ follow Gaussian distribution with 20% deviations accounting for the domain-to-domain variations of the parameters. As the number of domains increases from a single domain (1D) to a 12-domain (12D), the fluctuation range of current reduces from ~5 orders in 1D to around an order in 12D FeFET. The increment of domain number could suppress the overall dipole fluctuation as the result of the enhanced domain interaction with the increasing amount of domain, leading to the narrowing of the current fluctuation range. The variations of the stochasticity among different domain number FE p-bits lead to different probabilistic characteristics, notably the slope of the p-curves. For the comparison of the slopes, the p-curves are extracted with the same equivalent current threshold at $10^{-4}$ μA/μm, and the curves are further shifted to ensure at 0 V the probability of the output current is 50% in Fig. 6b. For the slope of the p-curves, determined by the stochasticity of the FeFET, it increases as the domain number changes from 1D to 12D, as the fluctuation range of the output current reduces.

**FeFET p-bits in the invertible logic circuit**

The integer factorization is implemented to verify the SC functions of the proposed FE p-bit, and we construct the IL circuits comprises of invertible AND gates (IAND) and Full Adders (FA) [2,34]. The probabilistic curves of FeFET p-bits, tuned by the gate voltage, are extracted to characterize the p-bits stochastic behaviors in circuit simulation. The overall circuit is first modelled through Verilog-A in HSPICE, with the connections of the invertible logic gates shown in Fig. 7a. An example of IAND connection is shown in Fig. 7(b), which is a passive resistor network following the connection rules mapping the correct solutions to the global energy minima (see Methods for circuit modeling details). The real-time operations of IAND are demonstrated in Supplementary Note 1. To perform the integer factorizations, a 2-bit×2-bit unsigned invertible logic circuit for the factorization of number 6 is modeled, with the solution probability shown in Fig. 7c. Terminal *A* and *B* are the output signals of the factorization, and the probabilistic nature of SC network produces constantly fluctuating results, corresponding to the multiple local energy minima in the IF network (Supplementary Material Fig. S1). The probability of each solution is further calculated by accounting the appearance of each solution within the operating time (up to 5 μs). The two correct solutions ($2 \times 3$ and $3 \times 2$) have the highest probability among other solutions, in accordance with the global energy minima configured in the circuit connections.

To further investigate the impact of FeFET's stochasticity on IF's performance, a 6-bit×6-bit IF circuit model is established in MATLAB for larger factorized digits. The p-bits characterized by different p-curves in Fig. 6b are implemented. For the factorization of digit 3233, the values, 53 and 61, are the correct results, and the probability of the correct solutions in $10^9$ cycles are summarized in Fig. 7d. It is worth noting that as the steepness of the p-curves increases, i.e., FE p-bits become less stochastic, the accuracy of the IL network enhances. The steeper slope of the p-curves facilitates the system's evolution to the energy minima, and hence, increases the portion of time the system stays in the correct solutions (global energy minima). Nevertheless, the further reduction of p-bits' stochasticity demonstrates the deterioration of the accuracy, as the probability for the IL network to escape from the local energy minima (wrong solutions) are reduced. This also widens the differences of probability between the two correct solutions (bar plots for 4,8, and 12D cases). Therefore, to optimize the performance for SC applications, the stochasticity of the p-bits, which is relating to the FE material parameters and domain dynamics, requires proper design and adjustment. Other circuit level optimization methods, such as the simulated annealing [35], could be implemented to improve the IF accuracy.

## Discussions

The domain-to-domain variation of Landau coefficients, domain interaction strength, and the amount of domain in FE systems are identified to play crucial roles in FE p-bits' stochasticity. In Fig. 6a, the fluctuation range of current, representing the stochasticity of p-bits, reduces from 4-5 orders to be less than an order (from 1D to 12D FeFET). It is worth noting that the suppressed current fluctuation range could render the thermal noise-induced probabilistic behaviors difficult to be differentiated from the other competing random processes and background noises in actual applications. Therefore, the feasibility of the proposed FeFET p-bits requires the proper design of its stochasticity. The recent miniaturization of HZO [36] and the two-dimensional FE [37-39] could provide a unique opportunity to exploit the advantages of the reduction of domain number to enhance the stochasticity FE system. Various material engineering methods, such as doping and defect engineering, could be utilized to increase the domain-to-domain variations in FE parameters, which could be implemented to widen the fluctuation range of polarizations.

Furthermore, it is noteworthy that the proposed FE p-bit has unique advantages comparing to the previous p-bits design (as compared in Table 1). Both FE-based and FM-based p-bits utilize the thermal noise-induced stochasticity with similar device density, significantly reducing the required area and energy consumption comparing to the full-CMOS counterparts. For the FM based p-bits, MTJ is originally a two-terminal device, requiring the connections with extra modules to achieve its p-bits functionality. The spin transfer torque (STT) based MTJ requires an extra FET to serve as the controlling terminals of p-bits [14], while the spin orbit torque (SOT) based MTJ [2] requires extra layer of heavy metal to separate the read-write paths. Comparing to the extra modules and circuit designs required in FM-based p-bits, the inherently three-terminal FeFET requires only a resistor for current-to-voltage conversions. The mature integration of FE layer in FeFET and CMOS-based circuits also facilitates the scaling of FE p-bits for large scale p-bits network in SC applications, comparing to the FM counterparts.

In summary, we propose a new p-bit design based on stochastic FeFET through thermal noise-induced polarization fluctuation. The mechanisms of the stochasticity in FE dipoles are investigated, and we reveal the key domain properties that play a crucial role in FE p-bits' stochasticity, including domain number, domain-to-domain variations, and domain interaction in the FE system. To verify the functionality of the FE p-bits, the integer factorization is performed based on the FE p-bits invertible logic circuit. The accuracy of the integer factorization is further found to depend on FE p-bits' stochasticity. The proposed FE p-bits excel in both extremely low hardware cost and compatibility with the CMOS technology node, rendering it a promising candidate in SC applications.

## Methods

**Phase-field model of the stochastic multidomain FE system**

The thermal noise-induced dipole fluctuations in the FE system could be characterized through a statistical mechanics [24-27,40] approach to capture the influence of the lattice vibrations in the time-dependent Landau-Ginzburg equations (TDGL), changing the equations into the stochastic form (Eq. 1). By considering the symmetry and orientation of the HZO, we assume only the out-of-plane component $p_y$ is non-zero, while its dynamics of the in-plane components are ignored (Fig. 1b). Moreover, with the assumption as the HZO is stress-free, the elastic strain energy density could be ignored in the FE system as well. As a result, for the enthalpy energy $W$, it includes the polarization potential energy $\Psi_{pol}$ up to six order, domain coupling energy $\Psi_{inter}$, free-space energy $\Psi_{el}$, external field $E_{fe}$ and the depolarization field $E_{dep}$ (Eq.2).

$$\mu \frac{\partial p(x,y,t)}{\partial t} = -\frac{\delta W}{\delta p(x,y,t)} + \nabla \cdot \frac{\partial W}{\partial \nabla p} + \mu \eta \qquad (1)$$

$$W(\varepsilon, e, p_y, \nabla p_y) = \Psi_{pol}(p_y) + \Psi_{inter}(\nabla p_y) + \Psi_{el} - E_{fe} \cdot p_y - E_{dep} \cdot p_y \qquad (2)$$

where $\mu$ represents the viscosity coefficient as $1.0 \times 10^1 \ \Omega \cdot m$, characterizing the deformation speed of the FE material. The term $\frac{\partial p(x,y,t)}{\partial t}$ shows the polarization transition rate (PTR) of the FE material. Each energy term in Eq. 2 as $\Psi_{pol}$, $\Psi_{inter}$, $\Psi_{el}$ and $E_{dep}$ could be expressed as:

$$\Psi_{pol}(p_y) = \alpha_1(T)p_y(t)^2 + \alpha_{11}p_y(t)^4 + \alpha_{111}p_y(t)^6 \qquad (3)$$

$$\Psi_{inter}(\nabla p_y) = \frac{G_0}{2}|\nabla p_y|^2 \qquad (4)$$

$$\Psi_{el} = -\frac{\varepsilon_0}{2}E_{fe}^2 \qquad (5)$$

$$E_{dep} = \frac{\lambda_1 + \lambda_2}{h\varepsilon_0 + (\lambda_1 + \lambda_2)\varepsilon_b}p_y \qquad (6)$$

where in Eq. 3, the $\alpha_1$, $\alpha_{11}$ and $\alpha_{111}$ are the parameters extracted from experimental measurements, and the first order approximation of $\alpha_1$ over temperature $T$ could be expressed as: $\alpha_1 = \alpha_0 \frac{T_c - T}{T_c}$, where the $T_c$ is the Curie temperature at $450 \ K$ [31]. In the simulations, it is assumed the entire FE system has uniform temperature $T$ at room temperature $(300 \ K)$. In Fig. 3, two different sets of Landau coefficients are implemented. For FE-1, the values are: $\alpha_0 = -4.0 \times 10^8 \ mF^{-1}$, $\alpha_{11} = 3.7 \times 10^9 \ m^5C^{-2}F^{-1}$, $\alpha_{111} = 1.1 \times 10^9 \ m^9C^{-4}F^{-1}$ [31]. For FE-2, the parameters are customized to possess higher energy barrier in free energy profile, and the values are: $\alpha_0 = -2.4 \times 10^9 \ mF^{-1}$, $\alpha_{11} = 1.3 \times 10^{10} \ m^5C^{-2}F^{-1}$, $\alpha_{111} = 1.1 \times 10^{10} \ m^9C^{-4}F^{-1}$. In single domain assumption with uniform polarization, the PTR's derivative against polarization could be approximated from Eq. 1 as:

$$\frac{\partial^2 p(x,y,t)}{\partial t \partial p} \sim -\frac{1}{\mu}\frac{\partial^2 \Psi_{pol}(p_y)}{\partial p^2} \qquad (7)$$

as the rest of the enthalpy energy terms are irrelevant to polarization, and the uniform polarization diminishes the energy aroused from spatially variating polarizations. For $\Psi_{inter}$ in Eq. 4, the coupling factor $G_0$ characterizes the coupling strength between different domains. Furthermore, to characterize the depolarization field in the system, the short-circuit boundary condition is assumed in 6 nm thick HZO material in a TiN/HZO/SiO$_2$ structure. The effective screening length $\lambda_1, \lambda_2$ normalized by $\varepsilon_0$ is 0.05 and 2.5 Å, respectively [31,41].

The thermal noise term $\eta$ in Eq. 1 is derived from the thermal noise-induced lattice vibrations and thus rendering fluctuation of the FE dipoles. The correlation function of the thermal noise could be expressed in Eq. 8 [24,40]. The overall distribution of thermal noise is assumed to follow the Gaussian distribution $G_{(0,1)}$ with the mean and deviations configured as 0 and 1 (Eq. 9). $V_{char}$ and $\Delta t$ represent the unit volume and time step in grid setting and time step in TDLG calculation, respectively:

$$< \eta(r,t), \eta(r',t') >_{t,\Omega} = \frac{2kT}{\mu} \delta(t-t')\delta(r-r') \tag{8}$$

$$\eta = \sqrt{\frac{2k_BT}{\mu V_{char}\Delta t}} \times G_{(0,1)} \tag{9}$$

The modeling aims to interpret the random noise term analogous to a random walk Brownian motion, and its overall probability, with sufficient observation time till equilibrium, would follow the Boltzmann-type probability distribution [24]. The overall TDLG equations in each discretized grids are solved by fourth-order Runge-Kutta methods in each discretized time step $\Delta t$, each grid is initialized to start from the spontaneous polarization opposite to the external applied electric field, in accordance with the pulse configuration demonstrated in the inset of Fig. 4a. The discretized time step is 0.02 ns, while the discretized volume $V_{char}$ is 1.38×10$^{-29}$ m$^3$, the unit cell size of HZO [42]. It is worth noting that the discretization of both volume and operation time might affect the impact of thermal noise [26]. Several discretization steps over volume and operation time have been tested in advance to ensure the discretization is properly configured. The robustness testing is further detailed in the Supplementary note 2.

In addition to the parameters discussed in the main text, some other parameters, such as the viscosity coefficient $\mu$ and temperature $T$, play important roles in characterizing the FE dynamic behaviors (in Eq. 1). Nevertheless, they either do not contribute to the polarization fluctuation (viscosity coefficient) or difficult to control in actual operations (temperature). Therefore, both factors' impact on FE stochasticity is further discussed in Supplementary note 3.

**Device modelling of the FeFET p-bits**

To model the dynamic behaviours of the stochastic FeFET with a 10 nm channel, the quasi-static conditions is assumed in each time step (at the scale of ns), as the response time of electron carriers are much faster than the deformation of the FE crystal structure [43-44]. The capacitance model is implemented to characterize the electrostatic relations of the TiN/HZO/SiO₂/Si structures (Fig. 1b). It could be expressed as:

$$Q_{FE} = Q_{sio_2} = Q_{MOS} \tag{10}$$

$$\left( \varepsilon_0 E_{FE} + \frac{1}{N} \sum_{n=1}^{N} P_n \right) = \frac{\varepsilon_0 \varepsilon_{sio2}}{t_{sio2}} V_{sio2} = \frac{\varepsilon_0 \varepsilon_{Si}}{t_{Si}} V_{MOS} \tag{11}$$

where the $Q_{FE}$, $Q_{sio_2}$, and $Q_{MOS}$ are the charge density of HZO layer, SiO₂ layer (2 nm thick) and the MOS structure. The serial connected capacitor conditions assumed in Eq. 10 could further be expanded in Eq. 11, where the $E_{FE}$ stands for the electric field in the FE layer, $P_n$ stands for the number of grids modelled in FE system, and its summation is normalized the total number of grids $N$. For SiO$_2$ layer and MOS structure, both are assumed to be infinitely large planar capacitors. For the channel potential, we assume the ideal gate control over the channel, and both source and drain harness no influence on channel potential [45]. This allows $V_{MOS}$ to directly represent the channel potential for carrier transport calculations.

For the carrier transport simulations, the combined schemes of Top-of-Barrier (ToB) [45] and Wentzel–Kramers–Brillouin (WKB) [46] schemes are implemented to model the ballistic and tunnelling current components in FeFET. The corresponding expressions are shown as follows:

$$I_{ToB} = \frac{q}{A} \iint \frac{d^2k}{4\pi^2} \, qv^+ f(E - E_{fs}) - \frac{q}{A} \iint \frac{d^2k}{4\pi^2} \, qv^- f(E - E_{fd}) \tag{12}$$

$$T(E, k_y) = \sum_m \exp\left(-2 \int_{X_{S,k_y}}^{X_{D,k_y}} \kappa_m(x)\right) dx \tag{13}$$

$$I_T = \frac{q}{2\pi h} \int T(E, k_y) \big[ f(E - E_{fs}) - f(E - E_{fd}) \big] dk_y dE \tag{14}$$

where $I_{ToB}$ and $I_T$ represent the ballistic and tunneling component of the total current, respectively. The ToB scheme in Eq. 12 details the summation of the forward and backward carrier states' occupation over the FBZ, as the $v^\pm$ represents the average group velocity in the forward/backward propagation states and $E_{fs}$, $E_{fd}$ stand for the fermi levels of source and drain. Both current components are normalized by the cross-section area of channel $A$. For the WKB scheme in Eq. 13, the transmission coefficient $T(E, k_y)$ is calculated, as $\kappa_m$ represents the decay rate of an electron in the $m$th band for a unit cell located at the $x$ position. Then, the total tunneling current is calculated through the Landau formula in Eq. 14.

**Invertible logic circuit modelling**

To demonstrate the feasibility of the FE p-bits' hardware implementation in SC applications, an invertible logic circuit consisting of FE p-bits is simulated in multiple levels. The circuit simulation flow of a specific invertible logic circuit is clarified as follows:

1) The essential characteristic of the p-bits is the sigmoidal response of the tunable probability with the external adjustment signals. The sigmoidal response of the typical p-bits [2] could be expressed as:

$$s_i(t) = sgn\{rand(-1,1) + \tanh[I_i(t)]\} \tag{15}$$

where $s_i$ is the binarized output states +1 and -1, while $I_i$ corresponds to the input of p-bit, and in our design, the gate bias of the FeFET. To obtain the sigmoidal relations from the FE p-bits, a dense gate bias sampling with the spacing of 0.1 V is implemented to the stochastic FeFET, with probability over the voltage threshold extracted in the operation time of 10 μs (~$10^5$ sampling time steps). Then, the sigmoidal fitting is performed to obtain the probabilistic curves of the corresponding FE p-bits (Fig. 1e). The extracted sigmoidal curve is then implemented in either the Verilog-A (used for HSPICE simulator in Virtuoso) or MATLAB (for MATLAB circuit simulation) in the format of the fitting curve or lookup table to characterize the probabilistic behaviors of the modelled FE p-bits circuit components.

2) The Linear Programming [34] is then implemented to obtain the coupling matrix $J$ (represents interactions among p-bits) and bias matrix $h$ (represents local bias to p-bits) with its truth table mapped to the lowest energy state of the network. For larger-scale invertible logic, like the invertible multiplier, several $J$ matrices and $h$ matrices of invertible AND gate, invertible Half adders, and Full adders need to be combined according to the corresponding logic circuit diagram.

For each p-bit in an invertible logic circuit, to calculate the input $I_i$ of the $i$th p-bit, the combined effects of the coupled p-bits components and local bias could be considered as:

$$I_i(t) = I_0\big[h_i(t) + \sum_j J_{ij} m_j(t)\big] \tag{16}$$

The $J$ and $h$ matrices are then translated into the electronic components i.e., the passive resistor network. During the circuits' operation, each p-bit is updated sequentially in every round of iteration to ensure the proper operations of the invertible logic networks [34]. The statistical results of each p-bits' state could be obtained with a sufficiently long observation time, aiming to approximate the probability under equilibrium conditions.

Following the characterizing schemes of the FE p-bits and circuit configurations, we simulate IL circuits comprising of various invertible logic gates. An invertible AND gate is first constructed with a circuit structure adapted from [2], as shown in Fig.7b. The passive resistor network is designed following the $J$ and $h$ matrices, with each resistor having a certain ratio over $R_0$ (500 $k\Omega$). The summed current signal at each node is further converted to voltage with current control voltage sources (CCVS) with the impedance of $10^6$ $\Omega$. The voltage of CCVS's input node is pinned at $V_{dd}$ /2 ($V_{dd}$ is at 0.8 V) to ensure the voltage input of each FE p-bit following the connection rules defined in $J$ and $h$ matrices. The real-time operations of the IAND are further demonstrated in Supplementary note 1. With the established invertible logic gate circuit models, the 2-bit × 2bit IF circuit is built following the logic gate connections shown in Fig. 7a. For integer factorization with larger numbers, the circuit modelling is performed in MATLAB, aiming to evaluate the impact of the FE p-bits' stochasticity. Following the characterizing scheme of FE p-bits with the assumption of ideal circuit components and connections, the accuracy of the IF operations under different probabilistic curves are evaluated, as shown in Fig. 7d.

## Data availability

The data that support the plots within this paper and the other findings of this study are available from the corresponding author upon reasonable request.

## Acknowledgements

This work at the National University of Singapore is supported by MOE-2017- T2-2-114, MOE-2019-T2-2-215, and FRC-A-8000194-01-00.

## Author contributions

L.G. and L.S. proposed the idea of FE p-bits project. L.S. conceived and performed the numerical simulations of the stochastic FE systems and analysed the data with L.G. G.X. and C.B. contribute to the FE properties and p-bit performance benchmark, respectively. H.Y. established the circuit models based on FE p-bits for integer factorization simulations. L.S. and L.G. prepared the manuscript with input from all other authors.

## Corresponding authors

Correspondence and requests for materials should be addressed to L.G. (Email: elelg@nus.edu.sg) and L.S. (Email: elelshe@nus.edu.sg).

## Competing interests

The authors declare no competing interests.

## References

[1] Shim, Yong, Akhilesh Jaiswal, and Kaushik Roy. "Ising computation based combinatorial optimization using spin-Hall effect (SHE) induced stochastic magnetization reversal." *Journal of applied physics* 121.19 (2017): 193902.

[2] Camsari, Kerem Yunus, et al. "Stochastic p-bits for invertible logic." Physical Review X 7.3 (2017): 031014.

[3] Faria, Rafatul, Kerem Y. Camsari, and Supriyo Datta. "Implementing Bayesian networks with embedded stochastic MRAM." *AIP Advances* 8.4 (2018): 045101.

[4] Alaghi, Armin, and John P. Hayes. "Survey of stochastic computing." ACM Transactions on Embedded computing systems (TECS) 12.2s (2013): 1-19.

[5] Gaines, Brian R. "Stochastic computing." *Proceedings of the April 18-20, 1967, spring joint computer conference*. 1967.

[6] Ribeiro, Sergio T. "Random-pulse machines." *IEEE Transactions on Electronic Computers* 3 (1967): 261-276.

[7] Qian, W., Li, X., Riedel, M. D., Bazargan, K., & Lilja, D. J. (2010). "An architecture for fault-tolerant computation with stochastic logic." *IEEE transactions on computers*, *60*(1), 93-105.

[8] Hammadou, T., Nilson, M., Bermak, A., and Ogunbona, P. 2003. "A 96 × 64 intelligent digital pixel array with extended binary stochastic arithmetic." In *Proceedings of the International Symposium on Circuits and Systems*. IV-772–IV-775.

[9] Dickson, J. A., McLeod, R. D., and Card, H. C. 1993. "Stochastic arithmetic implementations of neural networks with in-situ learning." In *Proceedings of the International Conference on Neural Networks*. 711–716.

[10] Liu, Yidong, Siting Liu, Yanzhi Wang, Fabrizio Lombardi, and Jie Han. "A survey of stochastic computing neural networks for machine learning applications." *IEEE Transactions on Neural Networks and Learning Systems* (2020).

[11] Pervaiz, Ahmed Zeeshan, et al. "Hardware emulation of stochastic p-bits for invertible logic." Scientific reports 7.1 (2017): 1-13.

[12] Onizawa, Naoya, et al. "A design framework for invertible logic." 2019 53rd Asilomar Conference on Signals, Systems, and Computers. IEEE, 2019.

[13] Kaiser, Jan, and Supriyo Datta. "Probabilistic computing with p-bits." *Applied Physics Letters* 119.15 (2021): 150503.

[14] Borders, W. A., Pervaiz, A. Z., Fukami, S., Camsari, K. Y., Ohno, H., & Datta, S. (2019). "Integer factorization using stochastic magnetic tunnel junctions." *Nature*, *573*(7774), 390-393.

[15] Mulaosmanovic, H., Mikolajick, T., & Slesazeck, S. (2017). "Random number generation based on ferroelectric switching." *IEEE Electron Device Letters*, *39*(1), 135-138.

[16] Tagantsev, Alexander K., Igor Stolichnov, Nava Setter, Jeffrey S. Cross, and Mineharu Tsukada. "Non-Kolmogorov-Avrami switching kinetics in ferroelectric thin films." Physical Review B66, no. 21 (2002): 214109.

[17] Gong, N., X. Sun, H. Jiang, K. S. Chang-Liao, Q. Xia, and T. P. Ma. "Nucleation limited switching (NLS) model for HfO2-based metal-ferroelectric-metal (MFM) capacitors: Switching kinetics and retention characteristics." *Applied Physics Letters* 112, no. 26 (2018): 262903.

[18] Alessandri, C., Pandey, P., Abusleme, A., & Seabaugh, A. (2019). "Monte Carlo simulation of switching dynamics in polycrystalline ferroelectric capacitors." *IEEE Transactions on Electron Devices*, *66*(8), 3527-3534.

[19] Luo, J., Liu, T., Fu, Z., Wei, X., Yang, M., Chen, L., Huang, Q. and Huang, R., 2021. "A Novel Ferroelectric FET-Based Adaptively-Stochastic Neuron for Stimulated-Annealing Based Optimizer With Ultra-Low Hardware Cost." *IEEE Electron Device Letters*, *43*(2), pp.308-311.

[20] Shur, Vladimir, Evgenii Rumyantsev, and Sergei Makarov. "Kinetics of phase transformations in real finite systems: Application to switching in ferroelectrics." *Journal of applied physics* 84.1 (1998): 445-451.

[21] Jo, J.Y., Yang, S.M., Kim, T., Lee, H.N., Yoon, J.G., Park, S., Jo, Y., Jung, M., Noh, T.W., 2009. "Nonlinear dynamics of domain-wall propagation in epitaxial ferroelectric thin films." Phys. Rev. Lett. 102, 045701.

[22] Puchberger, S., Soprunyuk, V., Schranz, W., Tröster, A., Roleder, K., Majchrowski, A., Carpenter, M.A. and Salje, E.K.H., 2017. "The noise of many needles: Jerky domain wall propagation in PbZrO3 and LaAlO3." *APL Materials*, *5*(4), p.046102.

[23] Sepliarsky, M., Phillpot, S.R., Streiffer, S.K., Stachiotti, M.G. and Migoni, R.L., 2001. "Polarization reversal in a perovskite ferroelectric by molecular-dynamics simulation." *Applied Physics Letters*, *79*(26), pp.4417-4419.

[24] Indergand, R., Vidyasagar, A., Nadkarni, N., & Kochmann, D. M. (2020). "A phase-field approach to studying the temperature-dependent ferroelectric response of bulk polycrystalline PZT." *Journal of the Mechanics and Physics of Solids*, *144*, 104098.

[25] Nambu, Shinji, and Djuniadi A. Sagala. "Domain formation and elastic long-range interaction in ferroelectric perovskites." *Physical Review B* 50.9 (1994): 5838.

[26] Wen, H., Liu, J., Chen, W., Xiong, W., & Zheng, Y. (2022). "Thermodynamics of polarization dynamics in ferroelectrics implemented by the phase field model." *Physical Review B*, *106*(2), 024111.

[27] Akamatsu, H., Yuan, Y., Stoica, V.A., Stone, G., Yang, T., Hong, Z., Lei, S., Zhu, Y., Haislmaier, R.C., Freeland, J.W. and Chen, L.Q., 2018. "Light-activated gigahertz ferroelectric domain dynamics." *Physical review letters*, *120*(9), p.096101.

[28] Liu, J., Wen, H., Chen, W. and Zheng, Y., 2021. "Atomistic studies of temporal characteristics of polarization relaxation in ferroelectrics." *Physical Review B*, *103*(1), p.014308.

[29] Grinberg, I., Shin, Y.H. and Rappe, A.M., 2009. "Molecular dynamics study of dielectric response in a relaxor ferroelectric." *Physical review letters*, *103*(19), p.197601.

[30] Liu, S., Grinberg, I. and Rappe, A.M., 2016. "Intrinsic ferroelectric switching from first principles." *Nature*, *534*(7607), pp.360-363.

[31] Hsu, C. S., Chang, S. C., Nikonov, D. E., Young, I. A., & Naeemi, A. (2020). "A Theoretical Study of Multidomain Ferroelectric Switching Dynamics With a Physics-Based SPICE Circuit Model for Phase-Field Simulations." *IEEE Transactions on Electron Devices*, *67*(7), 2952-2959.

[32] M. Hoffmann *et al.*, "Direct observation of negative capacitance in polycrystalline ferroelectric HfO2," *Adv. Funct. Mater.*, vol. 26, no. 47, pp. 8643–8649, Oct. 2016

[33] M. Hoffmann *et al.*, "Ferroelectric negative capacitance domain dynamics," *J. Appl. Phys.*, vol. 123, no. 18, May 2018

[34] Onizawa, N., Nishino, K., Smithson, S.C., Meyer, B.H., Gross, W.J., Yamagata, H., Fujita, H. and Hanyu, T., 2020. "A Design Framework for Invertible Logic." *IEEE transactions on computer-aided design of integrated circuits and systems: a publication of the IEEE Circuits and Systems Society*, pp.312-316.

[35] S. Kirkpatrick, C. D. Gelatt, and M. P. Vecchi, "Optimization by simulated annealing," *Science*, vol. 220, no. 4598, pp. 671–680, 1983.

[36] Cheema, Suraj S., Daewoong Kwon, Nirmaan Shanker, Roberto Dos Reis, Shang-Lin Hsu, Jun Xiao, Haigang Zhang et al. "Enhanced ferroelectricity in ultrathin films grown directly on silicon." *Nature* 580, no. 7804 (2020): 478-482.

[37] Fei, Zaiyao, Wenjin Zhao, Tauno A. Palomaki, Bosong Sun, Moira K. Miller, Zhiying Zhao, Jiaqiang Yan, Xiaodong Xu, and David H. Cobden. "Ferroelectric switching of a two-dimensional metal." Nature 560, no. 7718 (2018): 336-339.

[38] Xiao, Jun, Hanyu Zhu, Ying Wang, Wei Feng, Yunxia Hu, Arvind Dasgupta, Yimo Han et al. "Intrinsic two-dimensional ferroelectricity with dipole locking." Physical review letters 120, no. 22 (2018): 227601.

[39] Yuan, Shuoguo, Xin Luo, Hung Lit Chan, Chengcheng Xiao, Yawei Dai, Maohai Xie, and Jianhua Hao. "Room-temperature ferroelectricity in MoTe2 down to the atomic monolayer limit." Nature communications 10, no. 1 (2019): 1-6.

[40] Nambu, S., & Sagala, D. A. (1994). "Domain formation and elastic long-range interaction in ferroelectric perovskites." *Physical Review B*, *50*(9), 5838.

[41] Pal, H. S., Nikonov, D. E., Kim, R., & Lundstrom, M. S. (2012). "Electron-phonon scattering in planar mosfets: Negf and monte carlo methods." *arXiv preprint arXiv:1209.4878*.

[42] Materlik, R., Künneth, C. and Kersch, A., 2015. "The origin of ferroelectricity in Hf1−xZrxO2: A computational investigation and a surface energy model." *Journal of Applied Physics*, *117*(13), p.134109.

[43] Rollo, T., Wang, H., Han, G. and Esseni, D., 2018, December. "A simulation based study of NC-FETs design: Off-state versus on-state perspective." In *2018 IEEE International Electron Devices Meeting (IEDM)* (pp. 9-5). IEEE.

[44] Yan, M.H., Wu, M.H., Huang, H.H., Chen, Y.H., Chu, Y.H., Wu, T.L., Yeh, P.C., Wang, C.Y., Lin, Y.D., Su, J.W. and Tzeng, P.J., 2020, December. "BEOL-compatible multiple metal-ferroelectric-metal (m-MFM) FETs designed for low voltage (2.5 V), high density, and excellent reliability." In *2020 IEEE International Electron Devices Meeting (IEDM)* (pp. 4-6). IEEE.

[45] Rahman, A., Guo, J., Datta, S., & Lundstrom, M. S. (2003). "Theory of ballistic nanotransistors." *IEEE Transactions on Electron devices*, *50*(9), 1853-1864.

[46] Zhang, X., Lam, K. T., Low, K. L., Yeo, Y. C., & Liang, G. (2016). "Nanoscale FETs simulation based on full-complex-band structure and self-consistently solved atomic potential." *IEEE Transactions on Electron Devices*, *64*(1), 58-65.

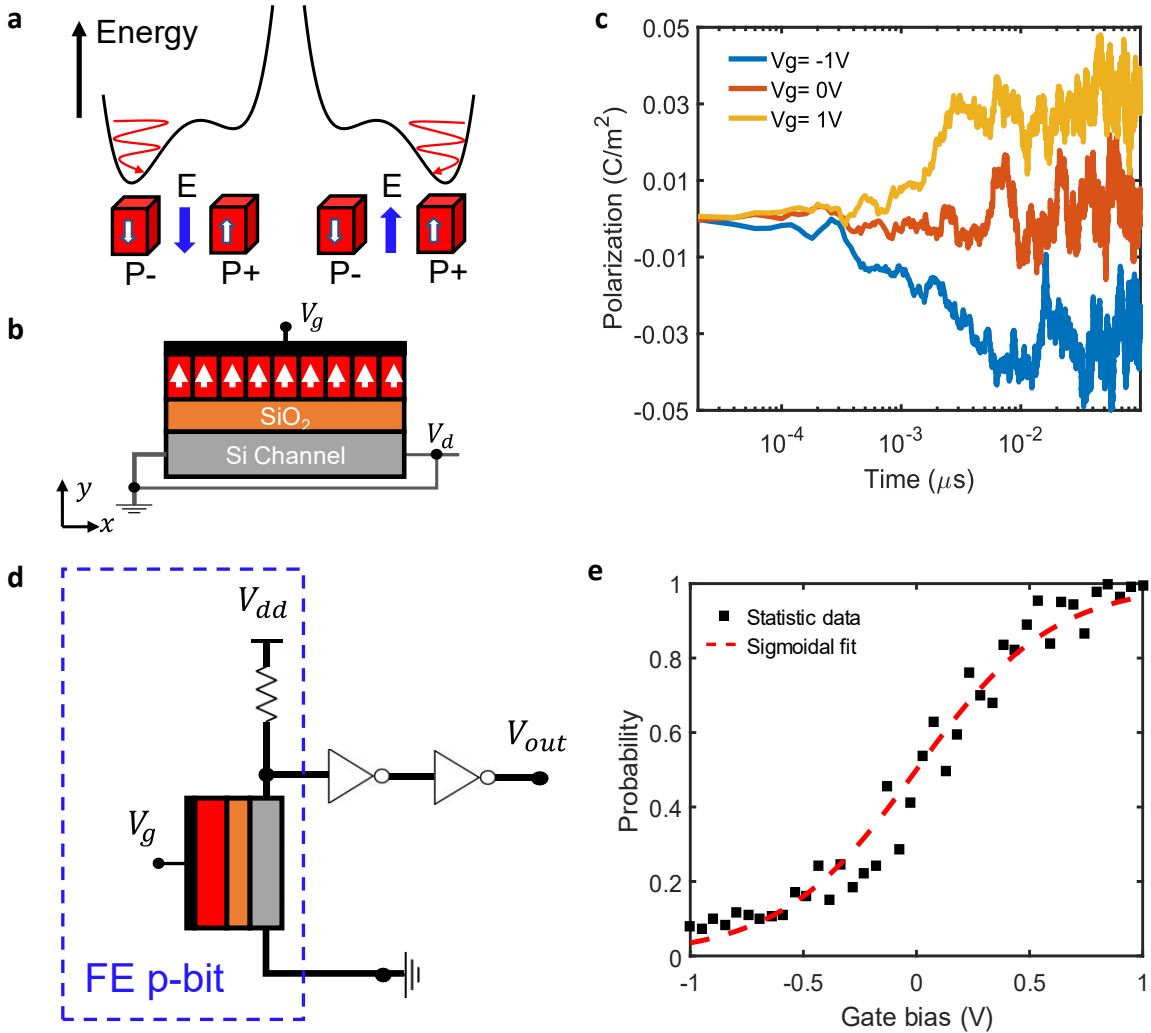

**Fig. 1 Overview of the FeFET p-bits. a** Thermal noise-induced polarization fluctuation in FE system. The free energy profiles of FE under two opposite electric field are shown. The chiral red arrows represent the fluctuating polarization around the spontaneous polarization. **b** Schematic view of the multidomain FeFET, red color area depicts the multidomain HZO layer. **c** Polarization fluctuations under different gate bias $V_g$. **d** The schematic view of the proposed FE p-bit and the post-processing modules. The stochastic FeFET produces the random analogue drain current, and the resistor converts the current to voltage signal. The inverters with voltage threshold $V_{th}$ further transform the analogue voltage signals to random binary bitstream $V_{out}$. **e** Probabilistic curve of the FE p-bits. The black dots are the extracted probability of $V_{out}$ surpassing the $V_{th}$, controlled by gate bias. The red line is the sigmoidal fit. The sigmoidal relation between probability and tuning signal (gate bias in this case) is the key characteristics of the p-bits for SC applications.

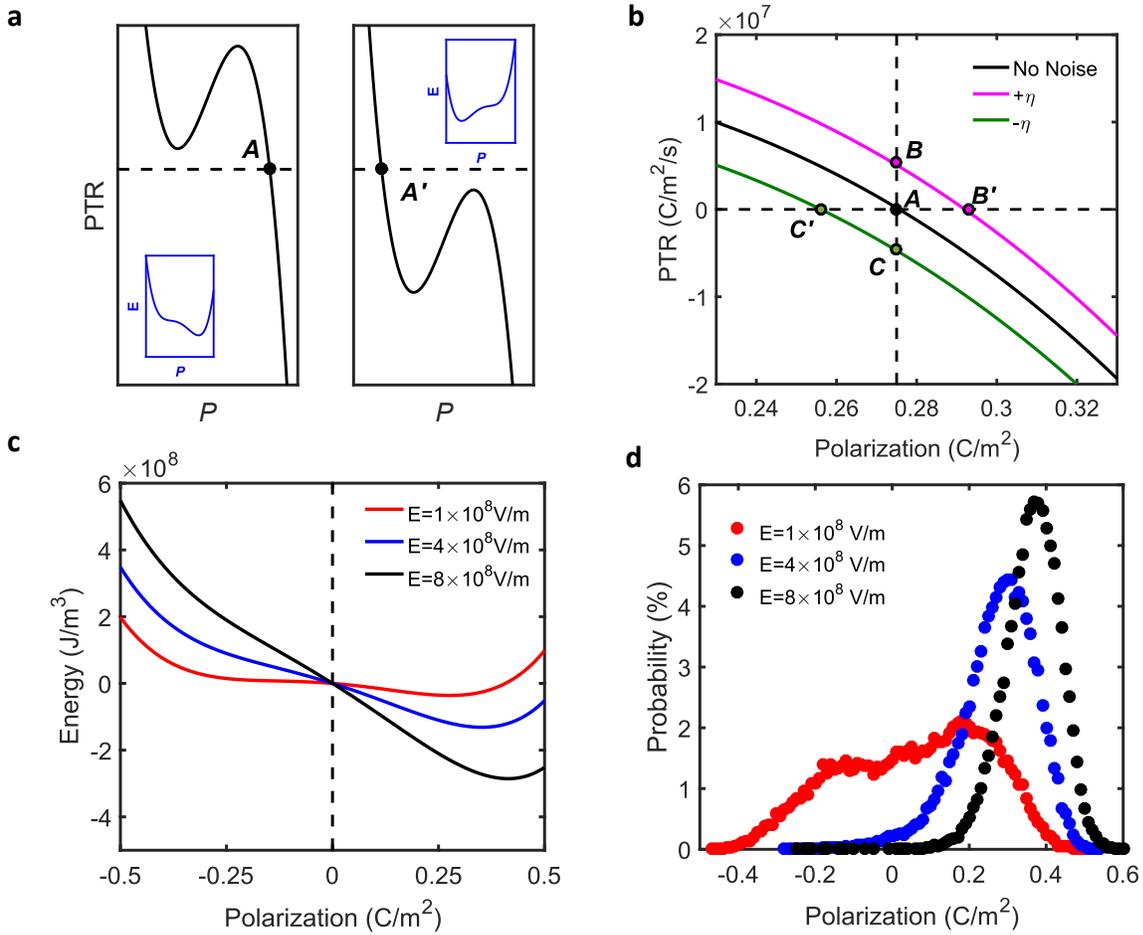

**Fig. 2 Mechanism of stochasticity in FE system. a** Polarization transition rate (PTR) of the FE system under the opposite electric fields. Insets are the corresponding polarization (P)-energy (E) profiles. Point *A* and *A'* represent the spontaneous polarization position. **b** Impact of thermal noise on FE system, as the thermal noise-induced lattice vibration $\pm\eta$ randomly facilitates/thwarts the polarization switching speed (from *A* to *B* or *C*). *B'* and *C'* represent the shifted SP positions due to the thermal noise. **c** Free energy profiles under various electric fields. The increasing external applied field reduces the energy of the SP. **d** Statistics of the polarization distribution within the observation time up to 1 µs (including $10^4$ sampling points) under different electric fields shown in **c**.

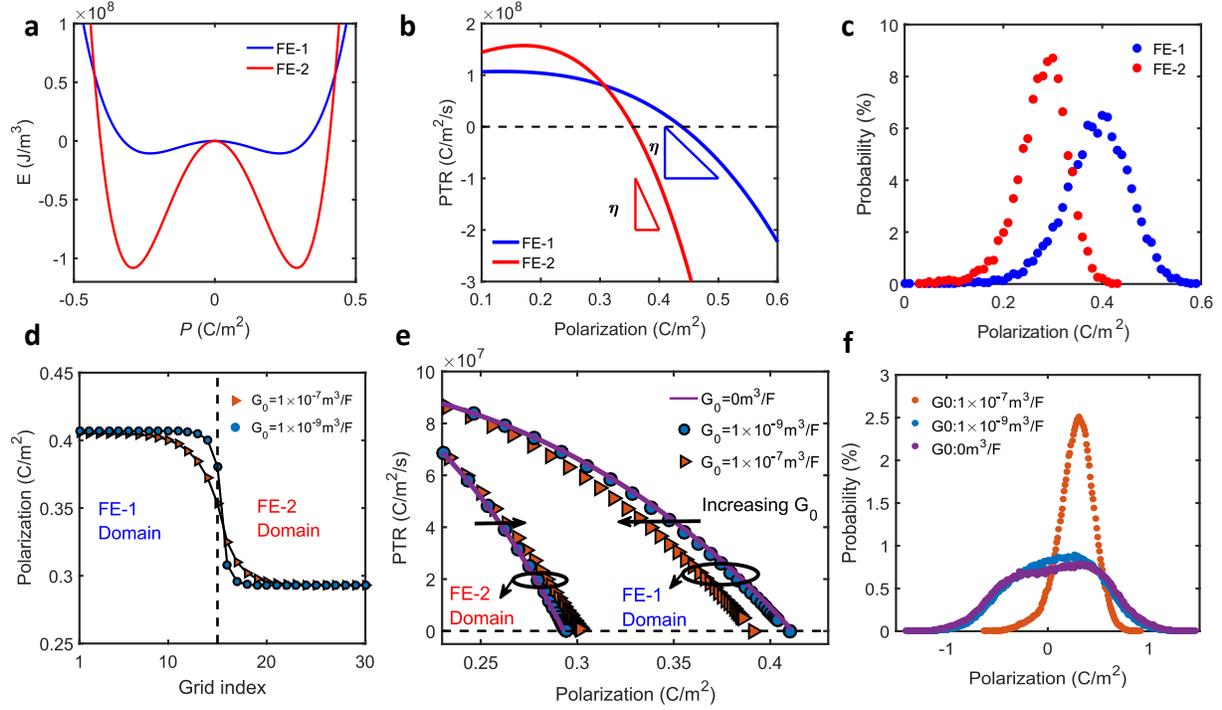

**Fig. 3 Impact of FE parameters on stochasticity. a** Free energy profiles of two different sets of Landau coefficients, FE-1 and FE-2. **b** The PTR of single domain system with FE-1 and FE-2 Landau coefficients under the electric field of $8 \times 10^8 \, V/m$. The red and blue triangles illustrate that under the same thermal noise factor $\eta$, the extent of the slopes of PTR could affect the range the polarization (the bottom lines of the triangles). **c** Statistic polarization distribution for single domain of FE-1 and FE-2. **d** Polarization profile of the two-domain system under different domain interaction parameter $G_0$ under the electric field $8 \times 10^8 \, V/m$. The domains have the FE-1 or FE-2 Landau coefficients, and the black dash line marks the boundary. **e** The PTR of the two-domain FE system under different domain interaction $G_0$. **f** Statistic polarization distribution under different domain interaction strength. The narrowed range of fluctuation corresponds to the reduction of SP variations among different domains in **e**.

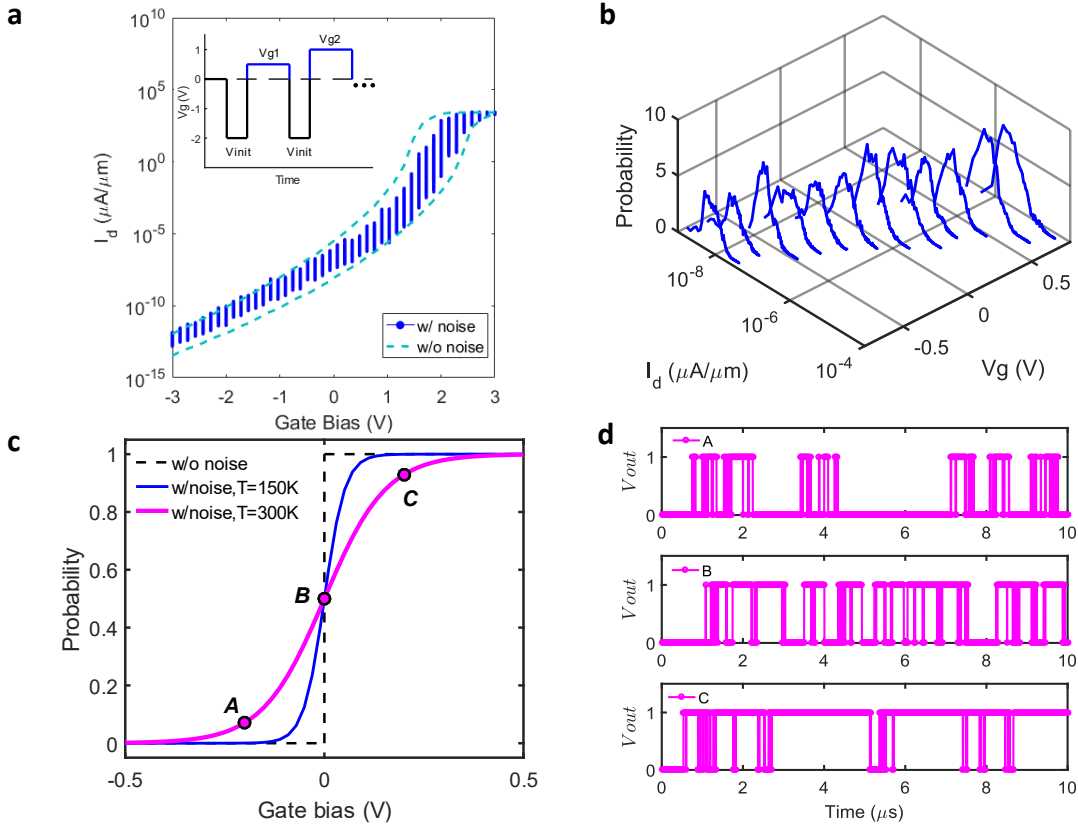

**Fig. 4 FE p-bits output signal and the extraction of probability curve. a** The *I-V* characteristics of the FeFET with FE system consisting of 12 domains within an observation time of 10 μs. The blue lines and cyan dash lines represent the FeFET w/ and w/o the impact of thermal noise. Inset: the configuration of gate bias $V_g$. Before each gate bias, the system is initialized through $V_{init}$ to enforce each domain starting from the same polarizations. Then, different gate bias ($V_{g1}$, $V_{g2}$, …) are applied to the system and remain constant within the observation time. **b** Zoom-in view of the current distribution in **a** of gate bias ranging from -1 to 1V. At each gate bias, the output current $I_d$ has different probability in the fluctuation. **c** Fitted probabilistic curve of p-bits with different stochasticity. Temperature *T* at 150 K and 300 K are implemented to adjust the impact of thermal noise, assuming uniform temperature within the system. *A*, *B* and *C* are the three sampled points to illustrate the extraction of the probability. **d** Different portions of time for p-bits surpasses the voltage threshold is demonstrated as gate bias at point *A*, *B* and *C*, with total sampling time up to 10 μs. $V_{out}$ is normalized by $V_{dd}/2$.

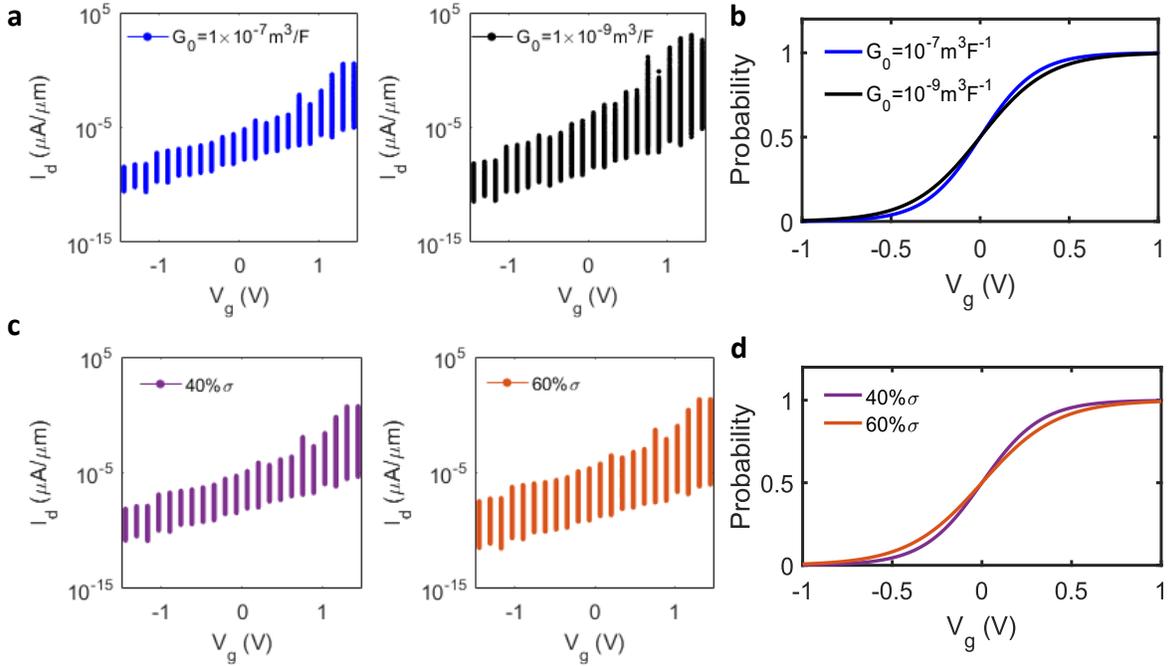

**Fig. 5 Impact of FE system's multidomain characteristics on p-bits stochasticity. a, b** *I-V* characteristics and probabilistic curves of FeFET consisting of four domains with different domain coupling strength. **c, d** *I-V* characteristics and probabilistic curves of FeFET consisting of four domains, with different percentage of variations in the Gaussian distribution of Landau coefficients and viscosity coefficients among the domains.

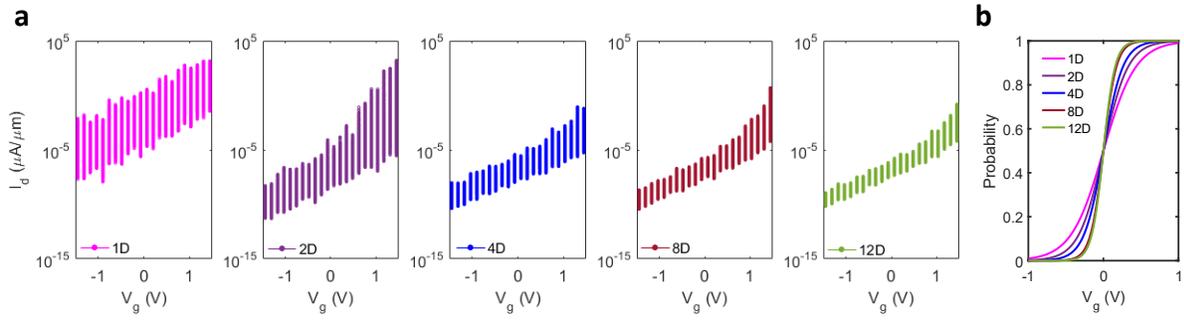

**Fig. 6 P-bits' stochasticity and the amount of domain in FE system. a** *I-V* characteristics of FeFET consisting of different number of domain (from single- to twelve-domain). The Landau coefficients and viscosity coefficients follow the Gaussian distribution of 20% σ among different domains. The uniform domain interaction $G_0 = 1 \times 10^{-7} m^3 F^{-1}$ is assumed for all cases. **b** The p-curves extracted from the corresponding *I-V* characteristics of FeFET with different number of domains in FE system.

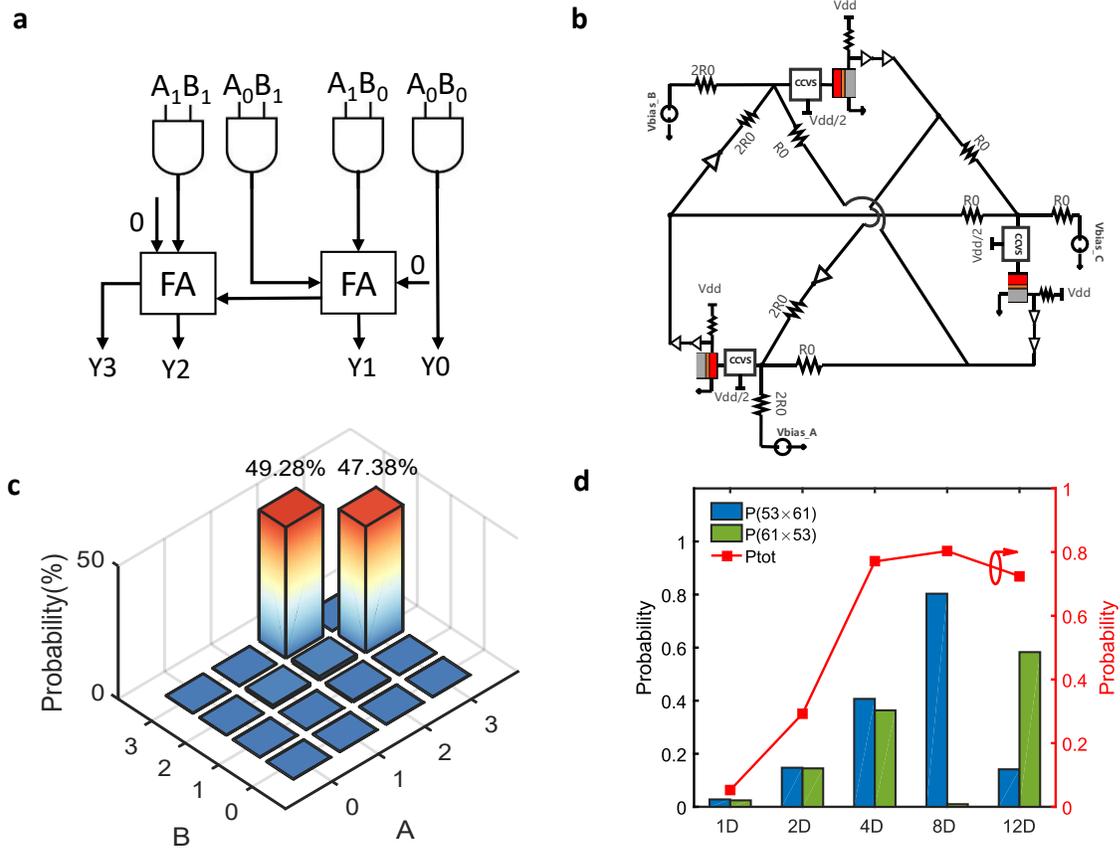

**Fig. 7 Integer factorization in invertible logic network based on FE p-bits.** **a** Invertible logic gate connections for 2-bit × 2bit integer factorizations. Invertible logic gates include both IAND and FA. *A* and *B* are the two output terminals in IF operation. **b** A circuit design example of IAND consisting of FE p-bits. The current-controlled voltage sources (CCVS) convert the proportionally summed current signal to voltage signal. $V_{dd}$ and $R_0$ are 0.8 V and 500 kΩ, respectively. **c** Probability of all solutions appeared in the IF operations. The correct solutions, 2×3 and 3×2, share the highest probability. **d** Bar/line plots of the probability of IF operations of 3233 with p-bits characterized by different p-curves from Fig. 6b. The Bar plots show the probability of each correct solution with different p-curves. The red line plots are the combined probability of obtaining the correct solutions.

**Table 1 | P-bits design comparison**

| Type | P-bits components | Source of stochasticity | CMOS Compatibility | Device Density |
|------|-------------------|-------------------------|--------------------|----------------|
| Fully-CMOS | LSFR+LUT [3] | Pseudo-RNG | Yes | ~ Few thousand (FETs) |
| | FPGA [13] | Pseudo-RNG | Yes | ~ Few thousand (FETs) |
| FM | Two-terminal MTJ [14] | Thermal noise | Complex | 3 (1$M$+1FET+1R) |
| | Three-terminal MTJ [2] | Thermal noise | Complex | 2 (1$M$+1$R$) |
| FE | FeFET (This work) | Thermal noise | Yes | 2 (1FeFET+1$R$) |

**Notes:** $M$ stands for MTJ, and $R$ stands for resistor.

# Supplementary Information

# Probabilistic-Bits based on Ferroelectric Field-Effect Transistors for Stochastic Computing

**Supplementary Note 1. Real time operations of integer factorization based on invertible logic circuit**

Based on the circuit network consisting of FE p-bits designed in Fig. 7, the integer factorization (IF) is performed in HSPICE through Verilog-A. As shown in Fig. 7a, the invertible logic (IL) circuit to perform the IF consists of invertible AND gates (IAND) and full adders. The IAND performs the reverse AND gate operations with the original output terminal C clamping at 0, while A and B are the output terminals in IAND. In Fig. S1a, the simulated results of IAND's IL networks are demonstrated, with output results constantly fluctuating among different solutions. The true table in Fig. S1b lists all the solutions of the IAND, with each solution modelled as the energy minima in IL circuit networks through circuit design. In Fig. S1c the probability of all the solutions is demonstrated, with three correct solutions sharing the highest probability (overall probability up to ~93%). The different probability among the correct solutions stems from the short observation time of the simulated system, as each correct solution has the same energy levels mapped into the IL networks. With a sufficiently long observation time, the probability of the correct solutions should be the same.

Furthermore, for the 2-bits by 2-bits IF operations, the output signals are demonstrated in Fig. S1d, as *A* and *B* are the output results for the factorization of 6. Similar to the IAND results, the constant fluctuation of output signals could be observed, with the probability of all solutions shown in Fig. 7c.

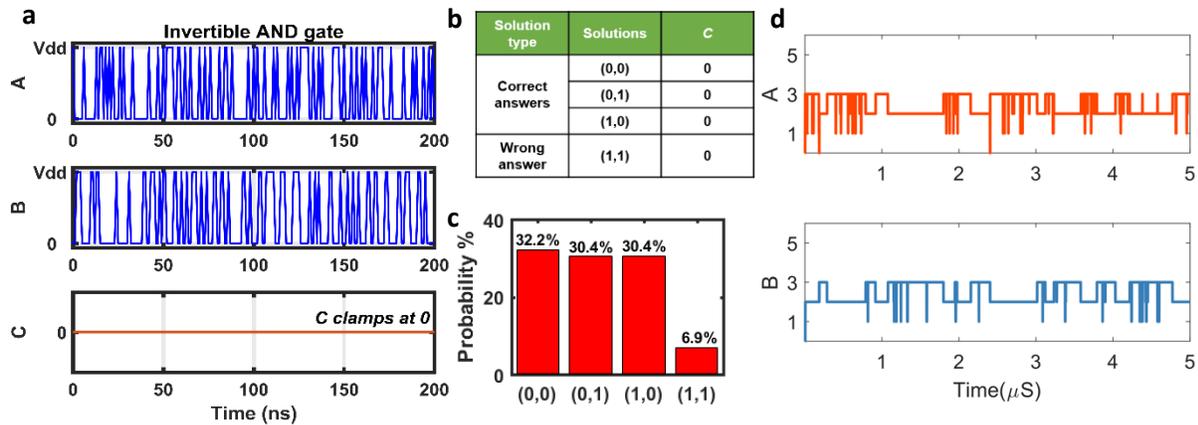

**Fig. S1 Real time output from the IAND and the 2-bits by 2-bits IF. a** Real-time output results IAND with terminal A and B as output and C clamps at 0. **b** True table of the IAND with C clamps at zero. **c** Probability of all solutions with C clamps at 0 under the observation time of 200 ns. **d** Real-time output results for 2-bits by 2-bits IF, with input clamps at 6 under the observation time of 5 μs.

**Supplementary Note 2. Discretization of operation time and system volume in TDGL**

For the solutions of the TDGL, the discretization of volume and time are necessary in the process. The discretized volume is usually configured as the size of the unit cell [1], which approximates the scale of the thermal noise-induced lattice vibration, such as the analysis of the phonon. The differences in the discretization of volume are demonstrated in Fig. S2a. $V_0$ is the original unit cell size of HZO implemented in the work, as $1.38 \times 10^{-28}$ m$^3$. Different volumes above/below $V_0$ represent looser/denser grids in the numerical simulations. It is worth noting that the further reduction of the grid suggests the overestimation of the lattice vibration, as it renders an increasingly wider range of fluctuations, coinciding with the results of several studies in phase field model [1-4].

For the discretization of time, the numerical solution of TDGL requires a sufficiently small-time step to ensure the convergence of the solutions. The impact of the time step in the dipole fluctuations is further demonstrated in Fig. S2b. Although the time steps vary for several orders, the overall polarization fluctuation range remains similar, demonstrating the stochasticity's robustness against the time step in thermal noise.

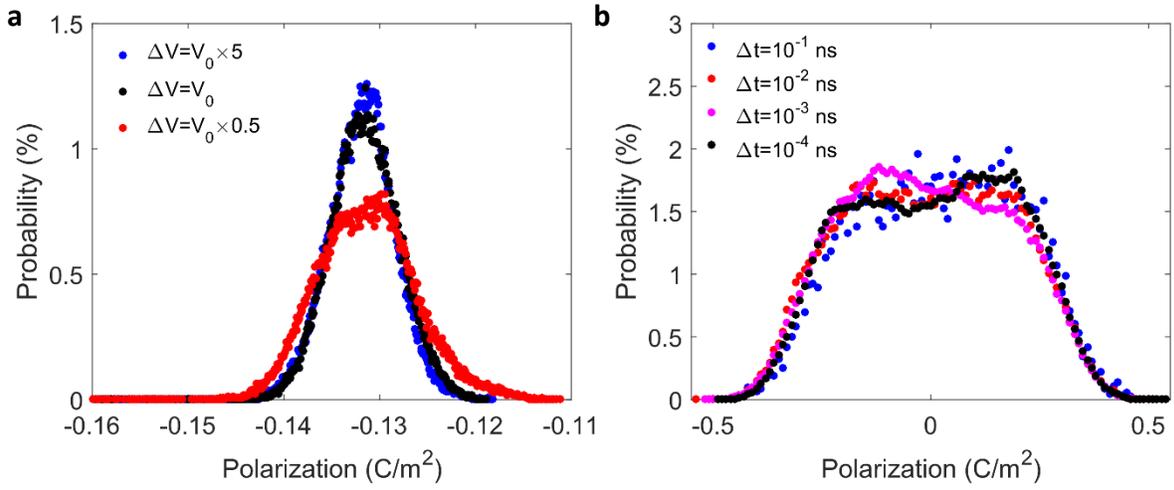

**Fig. S2 Distribution of polarization under time and volume discretization. a** Different volume discretized implemented in TDGL. $V_0$ is the unit cell volume of the HZO. **b** Distribution of polarization with different time steps.

**Supplementary Note 3. Impact of temperature and viscosity coefficients in FE stochasticity**

Both viscosity coefficients $\mu$ and temperature $T$ are directly involved in the FE dynamics. The thermal noise is inherent relating to the environment temperature, which is assumed to be uniform among the entire FE system in simulation. Besides, as the temperature relates to the amplitude of the lattice vibration, it directly characterizes the stochasticity of the FE system. The polarization dynamics under different temperatures are demonstrated in Fig. S3a, as the temperatures range from 100 to 300 K. The increasing temperature widens the range of dipole fluctuation, which is further demonstrated in Fig. S3b. The statistical distribution of polarization narrowed as the temperature reduces. It is worth noting that the center of the fluctuation is around the original spontaneous polarization position. The temperature only adjusts the range of fluctuation but does not shift the center of fluctuation like an externally applied electric field.

For viscosity coefficients $\mu$, the parameter is implemented to characterize the deformation speed of the FE material in the phase field model. In Fig. S3c, the polarization transition rate (PTR) is demonstrated with different $\mu$. By increasing $\mu$, under the same polarization, the PTR reduces, corresponding to the slower switching speed observed in the experiments [1]. It is worth noting that spontaneous polarization is not shifted under different $\mu$ (highlighted by the red dots). Although the viscosity coefficient does not directly affect the dipole fluctuations, it could still affect the FeFET-based p-bits in capacitance, as shown in Fig. S3d. The slope of the hysteresis loop represents the capacitance of the FE system, and the viscosity coefficient could induce different variations of capacitance between the polarization.

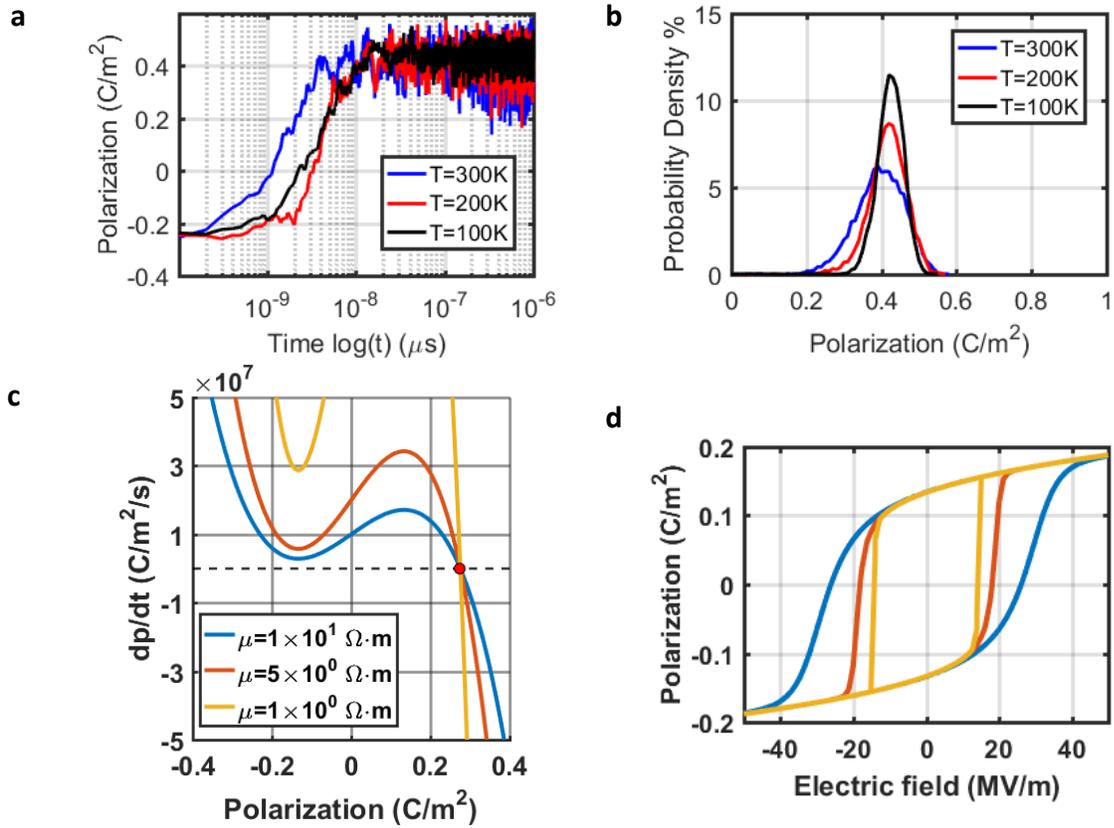

**Fig. S3 Impact of temperature and viscosity coefficients on FE stochasticity. a** Dynamic polarization fluctuation under different temperature. **b** Distribution of polarization under different temperature, with the observation time of 1 μs. **c** Polarization transition rate of the single domain FE under different viscosity coefficients. **d** Hysteresis curves under different various viscosity coefficients, demonstrating different slopes (capacitance) in *P-E* relation.

**References**

[1] Indergand, R., Vidyasagar, A., Nadkarni, N., & Kochmann, D. M. (2020). A phase-field approach to studying the temperature-dependent ferroelectric response of bulk polycrystalline PZT. *Journal of the Mechanics and Physics of Solids*, *144*, 104098.

[2] Nambu, Shinji, and Djuniadi A. Sagala. "Domain formation and elastic long-range interaction in ferroelectric perovskites." *Physical Review B* 50.9 (1994): 5838.

[3] Wen, H., Liu, J., Chen, W., Xiong, W., & Zheng, Y. (2022). Thermodynamics of polarization dynamics in ferroelectrics implemented by the phase field model. *Physical Review B*, *106*(2), 024111.

[4] Akamatsu, H., Yuan, Y., Stoica, V.A., Stone, G., Yang, T., Hong, Z., Lei, S., Zhu, Y., Haislmaier, R.C., Freeland, J.W. and Chen, L.Q., 2018. Light-activated gigahertz ferroelectric domain dynamics. *Physical review letters*, *120*(9), p.096101.